\documentclass[%
 reprint,
superscriptaddress,
 amsmath,amssymb,
 aps,
pra,
]{revtex4-1}

\usepackage{graphicx}
\graphicspath{ {/home/paromita/Downloads/arxiv/} }
\usepackage{dcolumn}
\usepackage{bm}

\begin{document}


\title{Studying the occupied and unoccupied electronic structure of LaCoO$_{3}$ by using DFT+embedded DMFT method with the calculated value of \textit{U}
}

\author{Paromita Dutta}
 \altaffiliation{dutta.paromita1@gmail.com}
 \affiliation{%
School of Basic Sciences, Indian Institute of Technology Mandi, Kamand, Himachal Pradesh-175005, India}%
 \author{Sohan Lal}%
\author{Sudhir K. Pandey} 
\affiliation{%
School of Engineering, Indian Institute of Technology Mandi, Kamand, Himachal Pradesh-175005, India
}%

\date{\today}

\begin{abstract}

In this work, we present a systematic study of the occupied and unoccupied electronic states of LaCoO$_{3}$ compound using DFT, DFT+\textit{U} and DFT+embedded DMFT methods. The value of \textit{U} used here is evaluated by using constrained DFT method and found to be $ \backsim $ 6.9 eV. It is found that DFT result has limitations with energy positions of PDOS peaks due to its inability of creating a hard gap although the DOS distribution appears to be fine with experimental attributes. The calculated value of \textit{U} is not an appropriate value for carrying out DFT+\textit{U} calculations as it has created an insulating gap of $ \backsim $ 1.8 eV with limitations in redistribution of DOS which is inconsistent with experimental spectral behaviour for the occupied states mainly. However, this value of \textit{U} is found to be an appropriate one for DFT+embedded DMFT method which creates a  gap of $\backsim $ 1.1 eV. The calculated PDOS of Co 3\textit{d}, La 5\textit{d}, La 4\textit{f} and O 2\textit{p} states are giving a remarkably good explanation for the occupied and unoccupied states of the experimental spectra in the energy range $\backsim $ -9.0 eV to $\backsim $ 12.0 eV.

\end{abstract}

\maketitle


\section{Introdution}

In condensed matter physics, understanding of strongly correlated electron systems is very crucial due to their unusual physical properties. These properties are extremely sensitive towards several parameters which give rise to many changes such as phase transitions, formation of complex patterns in chemically inhomogeneous situations etc., [1,2]. The firm understanding of these systems will build the platform for new applications to develop in near future. For the study of these systems' properties, some calculations are to be performed. These calculations require a very important parameter known as on-site Coulomb interaction \textit{U}. Generally, \textit{U} is taken as parameter, where one suitable value of \textit{U} is made to match with experimental results which gives a qualitative explanation of their physical properties. But this will be rather much more useful if a specific value of \textit{U} can be computed. This will be very specific to the material and can be used to study its properties in an extensive manner. This specific value of \textit{U} is termed as effective Coulomb interaction (\textit{U\textsubscript{eff}}) [3]. Though the physical meaning of \textit{U} is defined by Herring [4] as the energy cost for moving \textit{d} electrons between two atoms, where they have equal number of electrons [3]. Although \textit{U} can be related with unscreened Slater integral \textit{F\textsuperscript{0}}. However, in solids the \textit{U\textsubscript{eff}} is smaller than \textit{F\textsuperscript{0}} due to the screening effect present in them [3].

\textit{U} can be related to interaction between correlated orbitals. So, it is an important input parameter for performing the electronic structure calculations. For this normally two approaches are used for finding \textit{U}: (i). constrained density functional theory (cDFT) [3] and (ii). constrained random phase approximations (cRPA) [5-7]. cDFT computes \textit{U\textsubscript{eff}} value by varying energy from  number of correlated electrons on an impurity atom [3] while linear response theory is used by cRPA for obtaining the screened interaction [5-7]. Normally, it is found that the former one provides a larger \textit{U\textsubscript{eff}} value than the latter one due to the error occurring in compensating self-screening of localized electrons [6]. For example, the \textit{U\textsubscript{eff}} value for NiO is computed as $ \backsim $ 4.6 eV from cRPA [6] and $ \backsim $ 7.0-8.0 eV from cDFT [8-10]. And the value $ \backsim $ 5.0 eV goes with the experimental result [5]. Thus, cRPA is generally considered as a better approach than cDFT method. 

As in \textit{ab initio} calculations for strongly correlated electron systems, DFT has successfully been one of the useful tool in giving better explanations for their properties under study. Despite this, DFT has limitations in describing electronic and magnetic properties of many strongly correlated electron systems. For example, DFT could predict the magnetic properties of NiO-MnO series, but failed to explain the insulating ground state of CoO and FeO transition metal oxides [11,12]. This drawback was rectified by involving correlations effect. This correlation effect is applied to these systems by incorporating the parametric value of \textit{U} to perform electronic structure calculations. Corresponding results are then made to compare with experimental data to study the required properties of strongly correlated systems. This incorporation of Coulomb interaction can be achieved by two methods \textit{viz.} DFT+\textit{U} and DFT+DMFT [13-16]. In DFT+\textit{U} method, there is a static treatment of electronic correlations. Some properties of these systems have already been studied through DFT+\textit{U} method where normally \textit{U} is taken as parameter [13]. Moreover, it also is found that some of the experimental results cannot be well explained with this method [14]. So, more sophisticated approach  was needed something beyond DFT+\textit{U} and termed as DFT+DMFT [15-17]. In this method, DFT is used to obtain tight-binding model which corresponds to Wannier orbitals, then generalised Hubbard or Anderson type lattice model is solved within DMFT. In Millis \textit{et al.} work [17], it has been seen that several parameters needed to be tuned for predicting Mott gaps in early transition metal oxides. Later to have an equal footing in order to give better description of the itinerant and localised behaviour of correlated electrons, a stationary funtional (Luttinger-Ward functional [18]) with postulation of the locality of correlatios in real space [19] rather than in Wannier space was introduced. This later addition in second approach is termed as DFT+embedded DMFT functional (eDMFTF) by Haule \textit{et al.} [19]. In Haule \textit{et al.} work [19], it is showed that for a fixed value of \textit{U}, Mott gaps can be predicted accurately with experimental data. Thus, it is said that DFT+eDMFT gives a better result than DFT+DMFT due to combination of both the approximations [19,20].

Recently, Lal \textit{et al.} have used the self-consistently calculated \textit{U\textsubscript{eff}} from cDFT [3] approach to study the electronic structure of ZnV$_{2}$O$_{4}$ [21,22] compound by using DFT+eDMFT method. And the results of this study are in fairly good aggreement with the experimental results [23] for the occupied states. In the light of this result, it will be interesting to see whether the self-consistently  calculated value of \textit{U} from cDFT will be able to explain the electronic structure of other strongly correlated materials by using DFT+eDMFT method. As LaCoO$_{3}$ is the most studied material among strongly correlated systems and its spectra for the occupied and unoccupied states are available in literature [24,25]. It will be exciting to study its spectral properties by using  DFT+eDMFT approach. LaCoO$_{3}$ belongs to Perovskites family. This family forms an interesing class over many decades [2] and many of them falls into the category of strongly correlated electron systems. Due to their interesting physical properties such as charge ordering, phase transition, orbital ordering etc., they have been studied in large extent [26-33]. LaCoO$_{3}$ has unique electrical and magnetic properties due to the varying nature of spin state of Co\textsuperscript{+3} ion [26-33]. The crystallographic structure of LaCoO$_{3}$ is reported as rhombohedral having spacegroup as R-3c [34].

Here, we present a consistent study of the electronic structure calculations using DFT, DFT+\textit{U} and DFT+eDMFT approaches for LaCoO$_{3}$ compound for a calulated value of \textit{U} using cDFT method. The evaluated value of \textit{U\textsubscript{eff}} comes out to be $ \backsim $ 6.9 eV. The calculated DOS is made to compare with photoemission (PES) [24] and inverse photoemission spectroscopy (IPES) [25] experimental data. An attempt has been made to find out the extent of these methods' result in providing the explanation for the experimental spectral attributes of the occupied and unoccupied states.

\section{Computational Details}

The electronic structure calculations of LaCoO$_{3}$ has been carried out here. The calculations are divided into four parts as effective Coulomb interaction \textit{U\textsubscript{eff}}, DFT, DFT+\textit{U} and DFT+eDMFT. First three calculations are performed by the usage of full-potential linearized augmented plane-wave (FP-LAPW) method, where \textit{U\textsubscript{eff}} evaluation and the DFT part of the calculations are accomplished by WIEN2k code [35], DFT+\textit{U} part by Elk code [36] and the last part of the calculations are done within WIEN2k code[35] + code implemented by Haule \textit{et al.} [19]. For all these calculations, local density approximation (LDA) [37] is choosen as exchange-correlation. For crystal structural parameters (lattice parameters and atomic positions), a literature [38] is taken as reference where these parameters are observed experimentally. The muffin-tin sphere radii of 2.4, 1.9 and 1.6 bohr for La, Co and O sites, respectively with 8x8x8 k-point mesh size have been used for all calculatioins. 

For evaluating \textit{U\textsubscript{eff}} for Co 3\textit{d} atom in LaCoO$_{3}$, the spin-polarised calculation is carried out within constrained DFT. Anisimov \textit{et.al} [3] proposed a method for computing its numerical value. In this method, a hoping term (3\textit{d} orbital of one atom is connected with other obitals of remaining atoms) is constructed within a generalised supercell which is set to zero. \textit{U\textsubscript{eff}} for correlated 3\textit{d} shell is evaluated by the following formula, where the numbers of electrons are varied in non-hybridising 3\textit{d} shell.

\begin{eqnarray} 
\textit{U\textsubscript{eff}} = \epsilon_{3\textit{d}\uparrow}\Big(\frac{n+1}{2},\frac{n}{2}\Big)-\epsilon_{3\textit{d}\uparrow}\Big(\frac{n+1}{2},\frac{n}{2}-1\Big) \nonumber \\
 - \epsilon_{\textit{F}}\Big(\frac{n+1}{2},\frac{n}{2}\Big)+\epsilon_{\textit{F}}\Big(\frac{n+1}{2},\frac{n}{2}-1\Big)
\end{eqnarray}

where $\epsilon_{3\textit{d}\uparrow}$ and $\epsilon_{\textit{F}}$ are the spin up 3\textit{d} eigenvalue and the Fermi energy for n-up and n-down spins configuration, respectively with \textit{n} as total number of 3\textit{d} electrons. The implementation of procedure is done by Madsen \textit{et.al} [39], achieved through above mentioned code [35]. Then, the procedure for evaluating \textit{U\textsubscript{eff}} for LaCoO$_{3}$ is followed as given in Lal \textit{et al.} paper [22]. The evaluated \textit{U\textsubscript{eff}} comes out to be 6.9 eV.

For DFT part of each calculation, spin-unpolarised calculations are performed with same crystal parameters [38] as mentioned above. DFT+\textit{U} calculation as performed by using Elk code [36] with \textit{U\textsubscript{eff}} as evaluated. For this part of the calculation, the value of \textit{J} is calculated self-consistently within Elk code [36] as $ \backsim $ 1.25 eV. 
Here, DFT+\textit{U} calculation is carried out through Elk code [36] rather than WIEN2k code [35] due to the fact that in the former one, calculations perfomed are more generaleralised in terms of Hamiltonian consideration and moreover \textit{U} is taken as free parameter while \textit{J} is calculated self-consistently [40].

For DFT+eDMFT calculation, 1000 k-points grid size is used. This calculation is performed at room temperature where electronic charge density and impurity levels are self-consistent. Also to solve the auxilary impurity problem, a continuous-time quantum Monte Carlo impurity solver has been used here [41]. And the scheme for exact double-counting as proposed by Haule also has been used here [42]. For this calculation, Co \textit{t\textsubscript{2g}} orbitals are treated. With \textit{U} as 6.9 eV, $\lambda$ as 1.78 a.u\textsuperscript{-1} and \textit{J} as 1.18 eV, where these values are material specific. 
To obtain spectra on the real axis, the self-energy from the imaginary axis to real axis is met by using the analytical continuation. And for this analytical continuation maximum entropy method has been used [43]. For DOS calculation in DFT+eDMFT part, 2000 k-points grid is used.

\section{Results and Discussion}

In Fig. 1(a), the experimental LaCoO$_{3}$ spectrum, PES [24] (background subtracted) and IPES [25] data are shown. The partial density of states (PDOS) of La 5\textit{d}, La 4\textit{f}, Co 3\textit{d} and O 2\textit{p} states as obtained from DFT calculations are plotted in Fig. 1(b). These plots are divided into six  discrete regions \textit{viz.} I ($\backsim$ 10.0 eV onwards), II ($\backsim$ 3.3 eV to $\backsim$ 10.0 eV), III ($\backsim$ 0 eV to 3.3 eV), IV (-2.0 eV to $\backsim$ 0 eV), V ($\backsim$ -4.0 eV to $\backsim$ -2.0 eV) and VI ($\backsim$ -9.0 to $\backsim$ -4.0 eV) according to the features of spectra as observed experimentally. In order to compare the evaluated  DFT results with experimental spectra, we have divided both the plots of Fig. 1(a) and 1(b) into discrete regions within same energy range as mentioned. To identify the various contributions from different partial DOS for explaining the experimental spectra, we have noted some distinctive features and marked them in Fig. 1(a) as A, B, C, D, E and F observed $\backsim$ 8.7, $\backsim$ 6.7, $\backsim$ 2.0, $\backsim$ -1.0, $\backsim$ -3.0 and $\backsim$ -5.2 eV, respectively. Also for the better explanation of conduction band (CB) region, a linear background is added to the experimental spectra in Fig. 1(a).

\begin{figure}
  \begin{center}
    \includegraphics[width=0.73\linewidth, height=6cm]{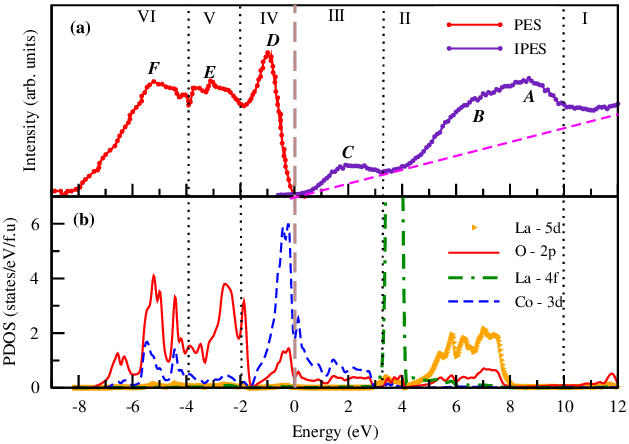}
  \end{center}

  \caption{\small{(a)Background subtracted photoemission spectroscopy measurements data [24] and the experimental inverse photoemission spectra [25] of LaCoO$_{3}$. The dashed line (Pink color) is drawn as linear background to the IPES data. (b)Partial density of states (PDOS) of La 5\textit{d}, La 4\textit{f}, Co 3\textit{p} and O 2\textit{p} states calculated within DFT. Zero energy corresponds to the Fermi level. }}
 
\end{figure}

It can be clearly seen from the Fig. 1(b) that there are finite PDOSs of Co 3\textit{d} and O 2\textit{p} at the Fermi level, indicating the metallic ground state which is exactly opposite to the experimentally observed insulating ground state. It tells that DFT is incapable of predicting insulating ground state for strongly correlated systems as it is expected from earlier works [44,45]. However, DFT is reported to provide a qualitative explanation for x-ray photoemmision spectra for the strongly correlated electron systems [46-48]. In the light of this, it will be interesting to see the extent of DFT in explaining the occupied and unoccupied states of experimental spectra for LaCoO$_{3}$ compound and the required study is given in the following paragraph.

On observing both the plots in Fig. 1(a) and 1(b), one can find that the energy positions of all peaks of PDOS are not matching with peaks of each experimental feature. Now when we start from region I, it is seen that the behaviour of DOS in the CB is  giving a similar behaviour to the experimental spectra. In this region I, the most contribution is coming from O 2\textit{p} states with an almost negligible contributions from rest of the states. In region II, the experimental spectra having a broader peak A and a hump marked as B. In this region, the electronic characters as shown by La 4\textit{f}, La 5\textit{d} and O 2\textit{p} having contributions as $ \backsim $ 67.0\%, $ \backsim $ 24.1\% and $ \backsim $ 8.1\%, respectively. From Fig. 1(a), for feature A it is expected to have DOS participations but it is observed that region around feature A in Fig. 1(b) is showing an almost zero DOS participation. Then, it is seen that the pattern followed by the peaks of La 5\textit{d} and O 2\textit{p} between $ \backsim $ 5.5 eV to $ \backsim $ 7.5 eV energies, is somewhat similar to the pattern shown by the feature B to feature A. Similarly, when we add both the PDOSs of La 5\textit{d} and O 2\textit{p}, the required pattern of the features will start to pursue by DOS here. But still, the last peak at $ \backsim $ 7.5 eV will still not be matching with peak A. Moreover, between the energy range $ \backsim $ 3.0 eV to $ \backsim $ 4.0 eV, the experimental spectra is expected to have zero DOS contribution after exclusion of linear background. But, we can see that our DFT calculated DOS is finite over here rather with dominant contribution from PDOS of La 4\textit{f}. And as we know that La 4\textit{f} cross-section is of about 10\% of La 5\textit{d} with 10.2 eV photon energy (source energy for IPES here) [49]. Thus, when we consider the cross-section of La 4\textit{f} with its DOS, then the height of its peak will be reduced but the peak will still be there in this energy range. Again, the issue is with energy positions of DOS here. This says DFT result in this region II, fails to provide the correct energy positions for La 5\textit{d} and O 2\textit{p} states although there is no problem with DOS distribution. So, it is suggesting of providing a constant rigid shift in this region II. Likewise, if a rigid shift of $ \backsim $ 1.0 eV is given to La 4\textit{f}, La 5\textit{d} and O 2\textit{p} DOS with consideration of their cross-sections. Then, the reason for the hump of B can be explained by suggesting that three peaks corresponding to La 4\textit{f}, La 5\textit{d} and O 2\textit{p} might be contributing. Thus, the whole pattern of hump B to broader peak A can be explained. Similarly, after rigid shift the peak of PDOS will start matching with peak A. Hence, with the addition of a constant rigid shift in region II, DFT can give a considerable explanation for spectral attributes here. Now in region III, there is a broader peak C. When we look at the region III, we find that there is one DOS peak $ \backsim $ 0.2 eV which is far away from peak C i.e., again energy position problem. Moreover, the pattern shown by PDOS of Co 3\textit{d} and O 2\textit{p} is not same as attribute shown by peak C. Thus, in this region III, DFT result has issues with both energy positions of states and DOS distribution. Now, even if we consider both the aspects of providing a rigid shift and taking into account the PDOS of Co 3\textit{d} and O 2\textit{p} with their cross-sections, still the pattern followed by feature C cannot be reached. Thus, for region III, DFT has failed to give even a qualitative explanation of experimental spectral attributes.

Now on comparing the valence band (VB) regions of Fig. 1(a) and 1(b), it is found that all the PDOS peaks against the experimental features D-F are not going with each other. 
Thus, the issue here again is with energy positions of calculated PDOS. Likewise, when we go to region IV it is observed that PDOS peak of Co 3\textit{d} is $\backsim$ 0.8 eV higher than peak D. Similarly, in region V, PDOS peak of O 2\textit{p} is $\backsim$ 0.5 eV higher than peak E while PDOS peaks appear to go with peak F in region VI. So, the peaks corresponding to PDOS in regions IV and V seem to be shifted towards the Fermi level. The behaviour shown by all the peaks of PDOS are reproducing the pattern as shown by all the experimental features, indicating an optimal DOS distribution by DFT. Thus, by giving a shift of $\backsim$ 0.8 eV and $\backsim$ 0.5 eV  to the regions IV and V, respectively, the similar behaviour of the attributes possessed by experimental features for the occupied states can be followed by calculated DFT. At the end, we can conclude that DFT is able to provide a reasonable account for the experimental attributes after giving an appropriate rigid shifts in different regions except for the region around peak C. This shift is needed beacause DFT is incapable of creating an insulating gap. Thus, an advanced method like DFT+\textit{U} may be useful as there will be a creation of the hard gap inherently. Owing to this, it will be interesting to know the extent of DFT+\textit{U} method in explaining the experimental spectra with the calculated value of \textit{U}. The study under DFT+\textit{U} is given in the following paragraph. 

 \begin{figure}
  \begin{center}
    \includegraphics[width=0.73\linewidth, height=6cm]{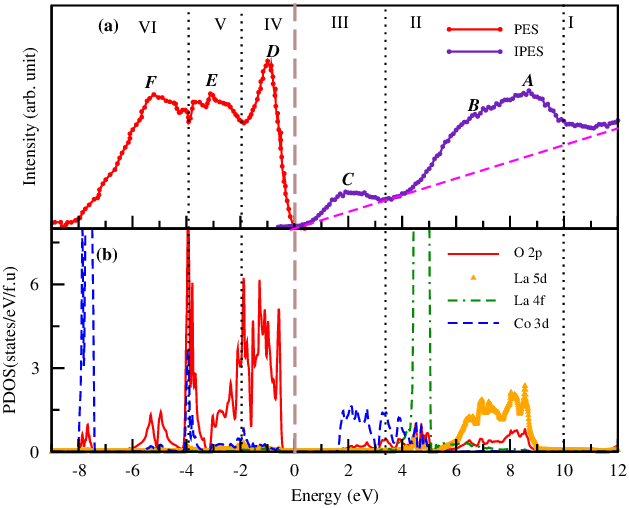}
  \end{center}

  \caption{\small{(a)Background subtracted photoemission spectroscopy measurements data [24] and the experimental inverse photoemission spectra [25] of LaCoO$_{3}$. The dashed line (Pink color) is drawn as linear background to the IPES data. (b)Partial density of states (PDOS) of La 5\textit{d}, La 4\textit{f}, Co 3\textit{p} and O 2\textit{p} states calculated within DFT+\textit{U} with \textit{U} is fixed at evaluated value of $ \backsim $ 6.9 eV. Zero energy corresponds to the Fermi level. }}
 
\end{figure}

In Fig. 2(b), the PDOS of La 5\textit{d}, La 4\textit{f}, Co 3\textit{d} and O 2\textit{p} states for as obtained from DFT+\textit{U} calculations are plotted. While looking at the plot, one can see that a large insulating gap of $\backsim$ 1.8 eV is created due to the inclusion of on-site Coulomb interaction. Although, correlation is applied to Co 3\textit{d} electrons, but one can perceive that it has modified all the energy peaks corresponding to different states with an enhancement in DOS distribution w.r.t DFT PDOS plot. Due to the creation of large gap, all the peaks in the CB and VB have shifted away from the Fermi level. 
Likewise, the peak of PDOS of La 5\textit{d} and O 2\textit{p} at 8.6 eV seems to match with A peak. As a result, DFT+\textit{U} is giving a reasonable explanation on the basis of both energy positions and DOS distribution for La 5\textit{d} and O 2\textit{p} atleast for peaks A and B. As for energy window $\backsim$ 3.3 eV to $\backsim$ 4.0 eV, it is expected to have zero DOS contributions. But, despite the negligible contribution from La 4\textit{f}, there is sudden enhancement of DOS contributions from Co 3\textit{d} and O 2\textit{p} due to the addition of correlations. Similarly, when we look at the PDOS peaks against feature C, it is noticed that more number of states has been generated due to the consideration of correlations again. On this account, the pattern followed by feature C cannot be followed by DFT+\textit{U} PDOS peaks. Now for the VB, it is seen that 
the correlation has affected the DOS of Co 3\textit{d} adversely. Like peak D is now showing contributions mainly from O 2\textit{p} states in contrary to DFT result. Similarly, there is a sharp peak of Co 3\textit{d} $\backsim$ -4.0 eV and $\backsim$ -7.8 eV which is absent in DFT plot. This shows that peak D has now only p character, indicating LaCoO$_{3}$ to be a charge transfer type insulator. This is in contrary to experimental observation [24], which says feature D has significant Co 3\textit{d} contribution. Mismatching of all the PDOS peaks with the experimental peaks can also be observed here. Based on the above discussion, one can say that the attributes of experimental features cannot be represented by PDOS peaks of DFT+\textit{U} except for peak A and the hump B. It is suggesting that DFT shows better results for spectral features than DFT+\textit{U} due to problem in redistribution of DOS within DFT+\textit{U}. It appears that the calculated value of \textit{U} from cDFT is an overestimated value for performing DFT+\textit{U} calculation w.r.t spectral features. There is a possiblity that a smaller value of \textit{U} might work for DFT+\textit{U} for the study of experimental spectra. As cRPA is known for giving smaller value of \textit{U} than cDFT. So, the calculated value of \textit{U} from cRPA may be an appropriate one to perform DFT+\textit{U} calculations. Hence, static mean field theory (DFT+\textit{U}) cannot provide a reasonable explanations for CB and VB of the experimental spectra with the calculated value of \textit{U} from cDFT. So, now it will be more appealing if one performs the dynamical mean field theory calculations for this compound with the calculated value of \textit{U}. Accordingly, we have calculated DOS for LaCoO$_{3}$ and the results are discussed in the following paragraph.
 
 \begin{figure}
  \begin{center}
    \includegraphics[width=0.73\linewidth, height=6cm]{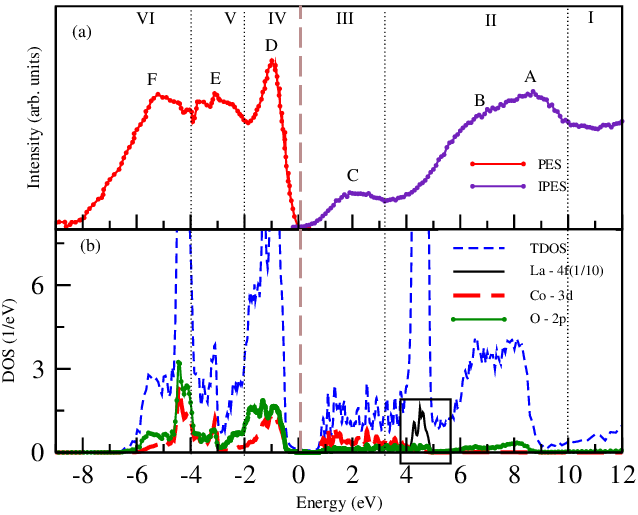}
  \end{center}

  \caption{\small{(a)Background subtracted photoemission spectroscopy measurements data [24] and the experimental inverse photoemission spectra [25] of LaCoO$_{3}$. (b)Total density of states (TDOS) calculated within DFT+eDMFT. And inside a square box, La 4\textit{f}$\times$(1/10) is drawn. Zero energy corresponds to the Fermi level. }}
  
\end{figure}

In Fig. 3(a)-(b), the spectra, total DOS, PDOS of Co 3\textit{d} and O 2\textit{p} states of LaCoO$_{3}$ within DFT+eDMFT calculations are shown. The creation of hard gap of $ \backsim $ 1.1 eV can be seen from the figure, indicating the insulating ground state for LaCoO$_{3}$ compound. This value of the gap is $ \backsim $ 0.7 eV less than the gap shown by DFT+\textit{U}. On looking at the CB part, one can find that all the PDOS peaks appear to go with all the experimental peaks when observed. For the energy range $ \backsim $ 6.5 eV to $ \backsim $ 8.5 eV of La 5\textit{d} and O 2\textit{p} are similar to the energy span for the hump B to peak A. Moreover, the attribute possessed by the hump B to peak A is similar to the line shape acquired by La 5\textit{d} and O 2\textit{p}. Now, as we know the cross-section of La 4\textit{f} is 1/10\textsuperscript{th} of La 5\textit{d} [49]. So, when the  DOS of La 4\textit{f} is reduced to 10\%, it starts appearing to have the similar behaviour of the rise of intensity $ \backsim $ 4.0 eV to the hump B (also shown in the square box as drawn inside the figure). Similarly, for peak C it is found that Co \textit{e\textsubscript{g}} states are contributing with same nature as peak C is showing in the figure. Thus, it shows that DFT+eDMFT is capable of representing the experimental CB spectra in better way than DFT and DFT+\textit{U}. For VB part, the line shape of all the PDOS peaks appear to be similar to the attribute shown by all the experimental peaks. However, in the region IV of VB, we can observe that the energy positions of PDOS peaks of Co 3\textit{d} and O 2\textit{p} are satisfying with the experimental peak D energy position. Although, for peak D, almost an equal contribution is coming from both the states of Co 3\textit{d} and O 2\textit{p} which is in contrary to DFT+\textit{U} result  and is more closer to the experimental result. Now on further going on the lower scale of energy, i.e., PDOS peaks against peak E seem to be similar with appropriate DOS distribution. However, for peak F all the PDOS peaks this time are somewhat shifted on the higher scale by $ \backsim $ 0.8 eV, i.e., energy position problem. 
Here, it is important to note that DFT+eDMFT result is able to provide a fairly good explanation for the experimental spectra except for bonding region i.e., for region around peak F while DFT results was showing discrepancy in all the calcuated DOS except for bonding region in the plot. Based on DFT+eDMFT calculations, the various contributions corresponding to distinct features can be given. For broader peak A and hump B, the maximum contribution is coming only from La 5\textit{d} and the region between $ \backsim $ 4.0 eV to $ \backsim $ 5.0 eV La 4\textit{f} is the only contributing one. And for peak C, Co \textit{e\textsubscript{g}} states are contributing. Similarly, for region around peak D almost equal contributions from Co 3\textit{d} and O 2\textit{p} states are coming. However, for peak E, O 2\textit{p} is showing the largest electronic character whereas for region around peak F, mixed contributions from both Co 3\textit{d} and O 2\textit{p} states are obeserved. In summing up, we can conclude that DFT+eDMFT is found to be capable of providing a quite reasonable account for the experimental CB and VB spectra both with the calculated value of \textit{U} using cDFT method. Owing to this work and from earlier work done by Lal \textit{et. al.} [21], it is tempting to suggest that the calculted value from cDFT is giving an appropriate value for \textit{U} for performing DFT+eDMFT calculations for explaining the experimental spectra. However, to establish this conjecture more works in this direction on different materials are required.

\section{Conclusion}

In this work, a consistent report on the electronic structure of LaCoO$_{3}$ by using DFT, DFT+\textit{U} and DFT+eDMFT approaches is provided. The value of \textit{U\textsubscript{eff}} as computed from cDFT method comes out to be as $ \backsim $ 6.9 eV. It is found that DFT has failed to create the hard gap due to which it has shown difficulties with energy positions of PDOS peaks in  going with the experimental spectral features while the DOS distribution seems to be similar to the experimental attributes. This says that DFT can explain the experimental features after providing an appropriate rigid shifts separately except for peak C in the CB. Although DFT+\textit{U} has created an insulating gap of $ \backsim $ 1.8 eV , still it has shown limitations in redistribution of DOS. Due to which DFT+\textit{U} cannot describe the experimental attributes except for peak A and the hump B in the CB. In consequence of this, it is suggesting that the value calculated from cDFT for \textit{U} is overvalued. However, from DFT+eDMFT calulations with the same value of \textit{U}, it is seen that DFT+eDMFT method has remarkably served the purpose i.e., band gap of $ \backsim $ 1.1 eV with reasonable explanation for the occupied and unoccupied electronic states of LaCoO$_{3}$ in the energy range $\backsim $ -9.0 eV to $\backsim $ 12.0 eV.

\section{Acknowledgements}

S.L is thankful to UGC, India for financial support.

\end{document}